\documentclass[a4paper,11pt]{article}
\pdfoutput=1 

\usepackage{jinstpub} 

\usepackage{subcaption}
\title{\boldmath Improving the time resolution of the MRPC detector using deep-learning algorithms}

\author[a]{F. Wang}
\author[a]{D. Han,}
\author[a,1]{Y. Wang\note{Corresponding author.}}

\affiliation[a]{Key Laboratory of Particle and Radiation Imaging, Ministry of Education, \\Department of Engineering Physics, Tsinghua University, \\Beijing 100084, China}

\emailAdd{yiwang@mail.tsinghua.edu.cn}

\abstract{
The multi-gap resistive plate chambers (MRPCs) will be used as the Time-of-Flight (ToF) system in the Solenoidal Large Intensity Device (SoLID). The time resolution required by the experiment for the MRPC system is 20 ps in order to make a 3 $\sigma$ separation of the $\pi/K$ created in the collisions. To achieve this goal, the whole system including the MRPC detector, the front-end electronics and the readout system will be upgraded. Based on the new system, a time reconstruction algorithm using a combined LSTM (ComLSTM) neural network is proposed. The best time resolution achieved with this algorithm in a cosmic ray test is 16.8 ps, which largely improves the timing ability of the MRPC detector and well satisfies the requirement of the SoLID.

}


\proceeding{XV$^{\text{th}}$ Workshop on Resistive Plate Chambers and Related Detectors\\
  10-14 February, 2020\\
  Rome, Italy}

\begin{document}
\maketitle
\flushbottom

\section{Introduction}
\label{sec:intro}
The multi-gap resistive plate chamber (MRPC) is a gaseous detector with parallel gap structures and has a really good time resolution. Over the years, it has already been used in many large physics experiments \cite{abelev_performance_2014,ackermann_star_2003,salabura_probing_2005,besiii_collaboration_construction_2009} and mostly as the Time-of-Flight (ToF) system. In the Jefferson National Lab (JLab), the Solenoidal Large Intensity Device (SoLID) plans to use the MRPCs as the ToF to perform the particle identification together with its heavy gas cerenkov detector. Since the beam energy is upgraded to 12 GeV, the $\pi$ and $K$ generated in the experiment are supposed to have momentum up to 7 GeV. Considering the flight distance of these particles, the ToF system should have a time resolution better than 20 ps so as to achieve a 3$\sigma$ separation of $\pi/K$ \cite{solid_collaboration_solid_2017}. However, the typical time resolution of the MRPC detectors currently used in large physics experiments is over 50 ps \cite{akindinov_final_2009,bonner_single_2003}, which is far from satisfactory for the SoLID. Therefore improving the time resolution of the detector system is one of the most important goals for its future development.

For the present MRPC detector system, the time uncertainty comes from 3 parts: detector, front-end electronics (FEE) and the readout system. To achieve the goal of 20 ps, we expect the  contribution of the time resolution from the detector to be below 15 ps, and the contribution from the FEE and the read out system to be around 10$\sim$15 ps. According to a previous study on the intrinsic time resolution of the MRPC \cite{wang_detailed_2020}, a detector with a time resolution of 15 ps can only be achieved when the gap thickness is below 0.16 mm and the number of gaps exceeds 20. In this work, two identical new thin-gap MRPCs are designed and produced. Each of them has 4 stacks and 8 gaps per stack, while the gap thickness is only 0.104 mm. These two detectors are amplified by a high performance FEE. The output of the electronics is the analog signal waveform of the detector which is then readout by a waveform digitizer. Comparing to the old electronics and the time-to-digital converter (TDC), the new system provides far more information about the induced signal and has a smaller time uncertainty.

Based on the new system, we proposed a time reconstruction algorithm using deep learning and neural networks. A combined long short term memory (ComLSTM) network which is an extension of our previous work \cite{wang_study_2020,wang_neural_2019} is designed and implemented to improve the MRPC time resolution from the perspective of the software. The simulation data used to train the network are optimized so that the most useful information is passed into the network and extracted by it. The time resolution of the MRPC given by the ComLSTM is 16.8 ps, which largely improves the timing ability of the MRPC detector and well satisfy the requirement of SoLID. ComLSTM can also be trained with the experiment data. In this work, two sets of the network methods based entirely on the experiment data are presented and their results are also at the scale of 20 ps.


\section{The time reconstruction algorithm and the ComLSTM network}
\label{sec:meth}
Deep neural networks have been proved to be powerful tools for solving highly non-linear pattern recognition problems. These kinds of algorithms have undergone tremendous innovations in the past 10 years and have already received wide attentions from the field of particle physics \cite{collaboration_neural_2014,aurisano_convolutional_2016}. Prior work that utilized the simple fully-connected (FC) or long short term memory (LSTM) network to reconstruct MRPC detection time has shown promising results, and therefore they are extended and improved in this work \cite{wang_study_2020,wang_neural_2019}.

\begin{figure}[htbp]
\centering
\includegraphics[width=.8\textwidth]{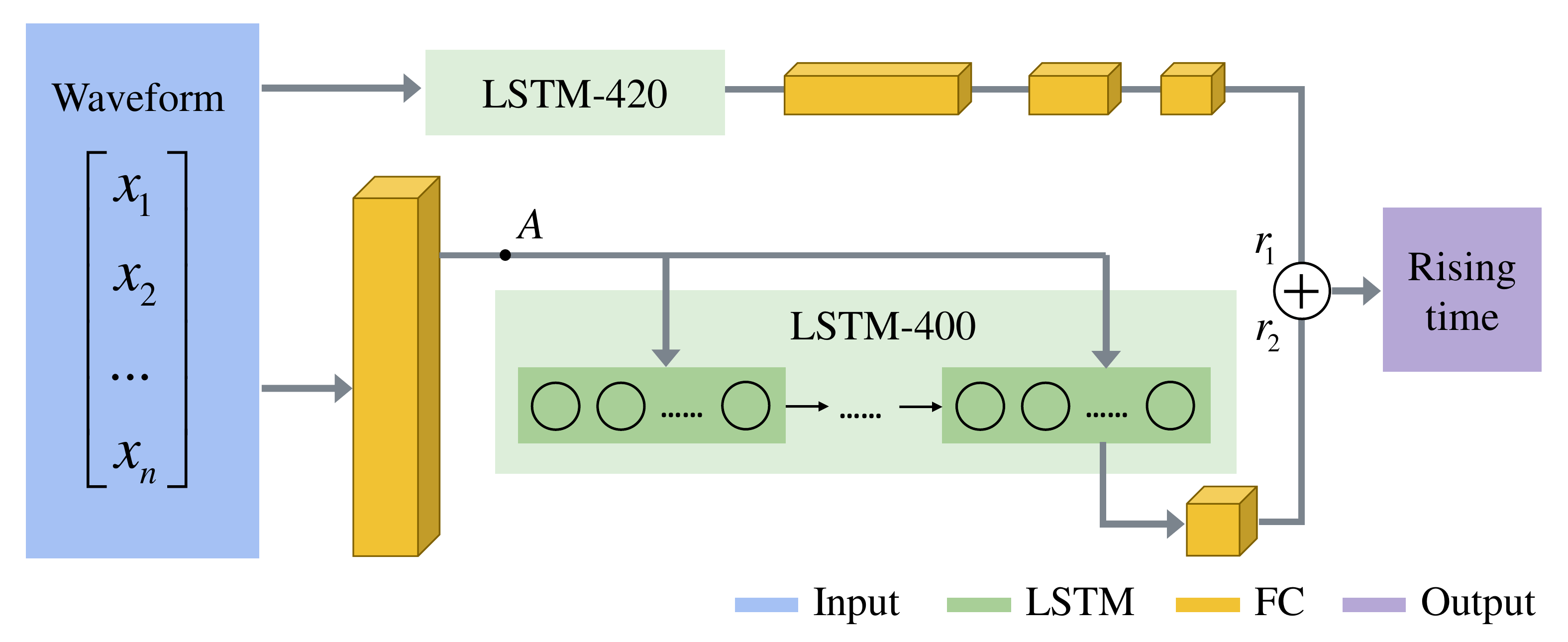}
\caption{\label{fig:comLSTM} The structure of the ComLSTM network.}
\end{figure}

LSTM is a special kind of recurrent neural network (RNN) that can effectively solve long sequential problems. An LSTM structure usually consists of many basic units, which are formed by a cell state, a hidden state and three different gates that control the relationship among the input, two states and the output. More details of the basic LSTM unit can be found in Ref \cite{wang_study_2020}. The ComLSTM (Combined LSTM) neural network proposed in this work combines the advantages of both the LSTM and FC. Its structure is shown in Figure \ref{fig:comLSTM}. The input $[x_1,x_2,...,x_n]$ is transmitted into two separate paths. Path $A$ connects an LSTM-420 and 3 fully connected layers with 128, 32 and 1 nodes respectively, while path $B$ is transmitted firstly into a fully connected layer with 100 nodes, then an LSTM-400 and finally a single FC layer with only 1 node. LSTM-$n$ represents a simple LSTM network, where the number of the time step is consistent with the dimension of its input vector, and at each time step there are $n$ basic units. The outputs of path $A$ and path $B$ are added together and weighted by two different coefficients $r_1$ and $r_2$ to adjust their importance for the final output. The activation functions for all the FC layers in ComLSTM is Rectified Linear Unit (ReLU) and the probability of dropout for regularization is 0.8. The loss function is the mean squared error between the true and estimated output of the network.

In the upgraded system of the SoLID MRPC, signals are readout as waveforms. If the incident particle arrives at the MRPC detector and interacts with the gas at time $t_0$, and the readout signal reaches the peak at $t_m$, then the rising time of the MRPC signal can be defined as $t_r=t_m-t_0$. The ComLSTM network works in the supervised manner. It takes several uniformly distributed points on the leading edge of the sampled signal as the input, and then extracts their embedding features and outputs the corresponding rising time $t_r$. Before feeding the data into the network, a peak searching algorithm should be applied to find $t_m$. The training data of the network is from a Monte Carlo simulation of the detector working in the same condition as the experiment using the framework developed in our group \cite{wang_standalone_2018}, because the true $t_0$ and thus $t_r$ is known in the simulation. The network is tested with the experiment data. If the simulation signals are consistent with the experiment, then the information extracted from the simulation is useful for reconstructing the time for the experiment and thus an accurate estimation of $t_r$ which also means $t_0$ ($=t_r-t_m$) can be obtained.

\section{Results}
\label{sec:res}

\subsection{Experiment}
\label{sec:exper}
To achieve the goal of 20 ps time resolution and prove the effectiveness of the ComLSTM, two identical MRPCs are produced and tested with the cosmic rays. Each of these MRPCs has 4 stacks and 8 0.104 mm gaps per stack. The resistive plates are made of the floating glass with a thickness of 0.5 mm, and the read out strips on the PCB (Printed Circuit Board) are 7 mm wide (3 mm interval).

\begin{figure}[htbp]
\centering
\includegraphics[width=.48\textwidth]{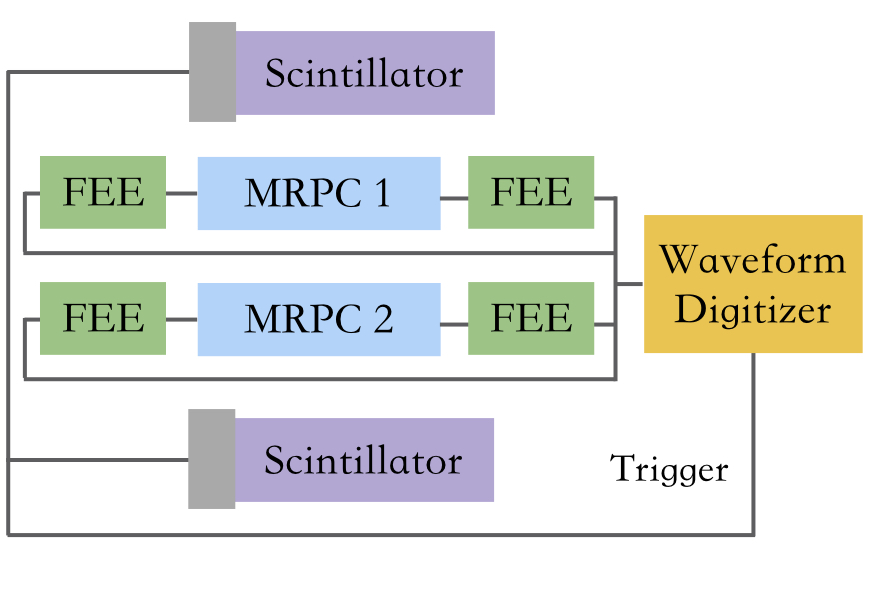}
\caption{\label{fig:setup} The setup of the cosmic ray experiment.}
\end{figure}

The setup of the cosmic ray experiment is shown in Figure \ref{fig:setup}. Two MRPCs are placed one on top of the other, and they are read out from both sides of the PCB strip. The coincident events of two scintillators above and below the MRPCs provide triggers for the system. The induced differential signals of MRPCs are amplified by a high performance front-end electronics (FEE) with a bandwidth of 1.3 GHz and read out by a Lecroy HDO6104A oscilloscope which has a bandwidth of 1 GHz and a sampling rate of up to 10 Gs/s. The rising time of a typical MRPC signal collected in the oscilloscope is around 1 ns and around 10 sample points are recorded on the leading edge. 

\subsection{Data processing} 
\label{sec:data}
Due to the channel limitation of the oscilloscope, only 1 strip (left+right) per MRPC can be recorded. Therefore, events with a signal-to-noise ratio of less than 20 are discarded to ensure that the incident position of the selected particle is in the readout strip area. Vertical selection that controls the signal peak time difference of two MRPCs on the left (right) strip is also made to filter out the non-vertical cosmic rays.

As the neural network is trained with simulation data and tested with the experiment, the consistency of these two datasets is crucial to the success of the algorithm. In the simulation, parameters of the FEE relates much to the shape of the signal and are adjusted according to the experiment. In order to quantitatively evaluate the differences between the simulation and the experiment, the Kullback-Leibler (KL) divergence which is originally used to describe the differences between two probability distributions $p(x)$ and $q(x)$ is introduced. For discrete variables, it can be defined as:
\begin{equation}
\label{eq:dpq}
D(p,q)= \sum_{x\in X}{p(x)\frac{p(x)}{q(x)}}
\end{equation}

The non-negative KL divergence $D(p,q)$ represents the information lost when using an approximating distribution $q(x)$ to estimate the true distribution $p(x)$. The greater the differences between $p$ and $q$, the greater the $D(p,q)$, and $D(p,q)=0$ if and only if $p(x)=q(x)$. In the case of MRPC, as the signals of both the experiment and simulation are digitized by a digitizer with a sampling rate of 10 Gs/s, then around 10 points are recorded on the leading edge. If the peak point on every signal is defined as point 10, and the points before it is defined in order as points 9,8...,1, then the amplitude distribution of every point along the leading edge is regarded as $p_i(x)$ (i=1,2...10), while the amplitude distribution of the corresponding point on the simulation signal is $q_i(x)$. The detector is simulated over 100 times with different sets of the FEE parameters, and if the $D_i^k(p_i,q_i^k)$ is defined to be the KL divergence for point $i$ in simulation dataset $k$, then the KL divergence between the experiment and this simulation is:
\begin{equation}
\label{eq:avedpq}
D^{k}(p,q)= \frac{1}{10}\sum_{i=1}^{10}{D_i^k(p_i,q_i^k)}
\end{equation}

The ComLSTM neural network is trained separately with these simulation datasets. The input of the network is 10 uniformly distributed points on the leading edge, numbered from 1 to 10. The initial learning rate is 0.001 and decreases during the training. All the networks are converged after about 100-200 epochs, and the models are used to predict the rising time of the waveforms collected on the left and right sides of the upper and lower MRPCs, which are $t_{l1}$, $t_{r1}$, $t_{l2}$, $t_{r2}$. To evaluate the time resolution, the time difference between the two MRPCs is defined as:
\begin{equation}
\label{eq:deltat}
\Delta t=\frac{t_{l1}+t_{r1}}{2}-\frac{t_{l2}+t_{r2}}{2}
\end{equation}
Since two MRPCs are produced and work independently, the time resolution of the detector is:
\begin{equation}
\label{eq:sigmat}
\sigma_t=\frac{\sigma(\Delta t)}{\sqrt{2}}
\end{equation}

\begin{figure}[htbp]
\centering
\includegraphics[width=.5\textwidth]{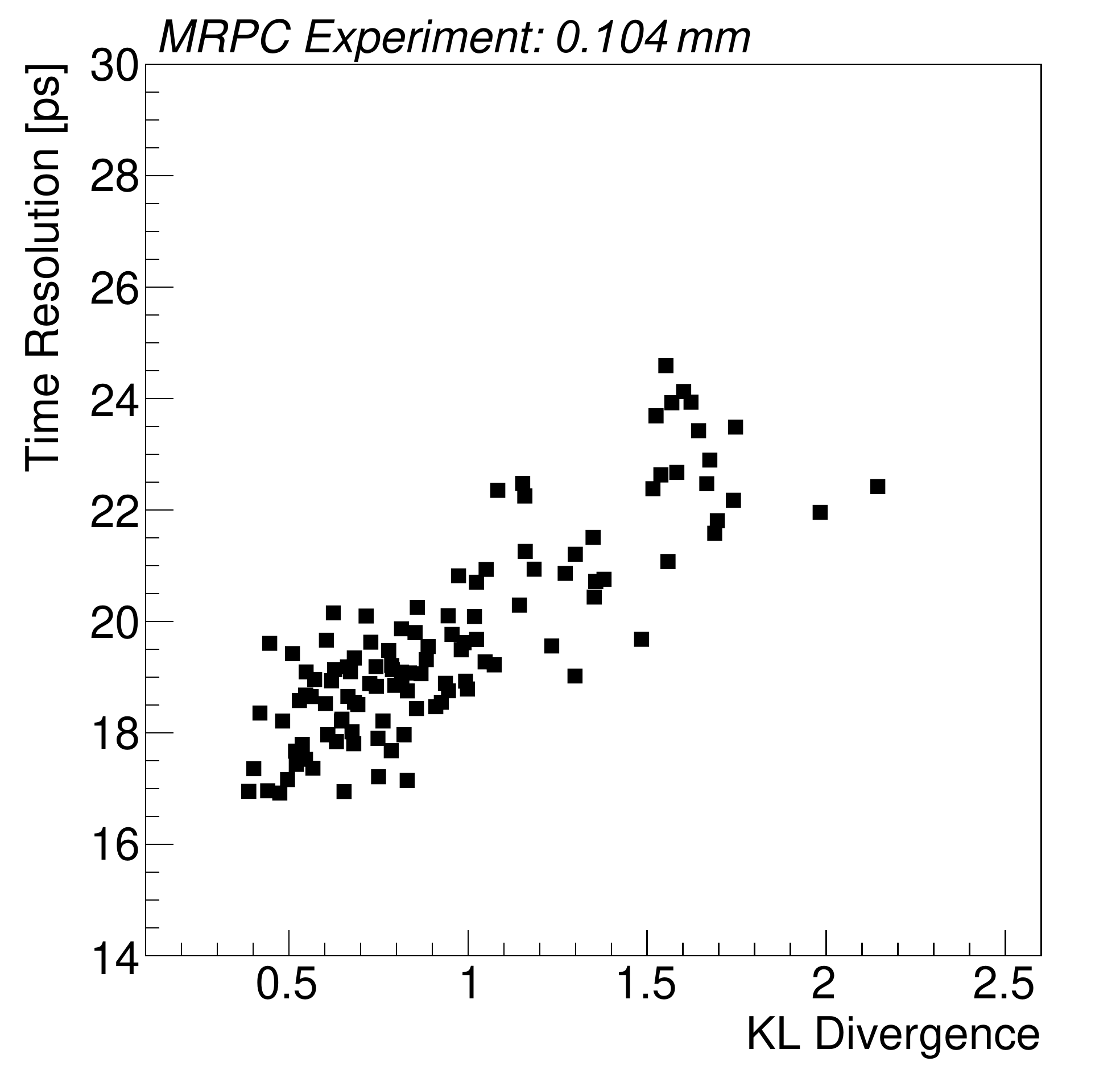}
\caption{\label{fig:kl} The relationship between the time resolution and the KL divergence.}
\end{figure}

Figure \ref{fig:kl} shows the relationship between the KL divergence and the time resolution for different simulation datasets when MRPCs work at 156 kV/cm. It is clear that the resolution increases almost linearly with the KL divergence, which means if the correlation between the experiment and the simulation used to trained the network is stronger, the performance of the neural network will become better. KL divergence is a good measurement of the waveform similarity, and therefore the parameters of the simulation with a KL divergence of only 0.39 is chosen to generate the training data for all the network in the following parts.

\subsection{Detector performance}
\label{sec:perform}

MRPCs in the experiments are tested at different electric field and their timing performances are analyzed with both the ComLSTM neural network and the traditional threshold based method. 

\begin{figure}
    \centering
    \begin{subfigure}[b]{0.48\textwidth}
        \includegraphics[width=\textwidth]{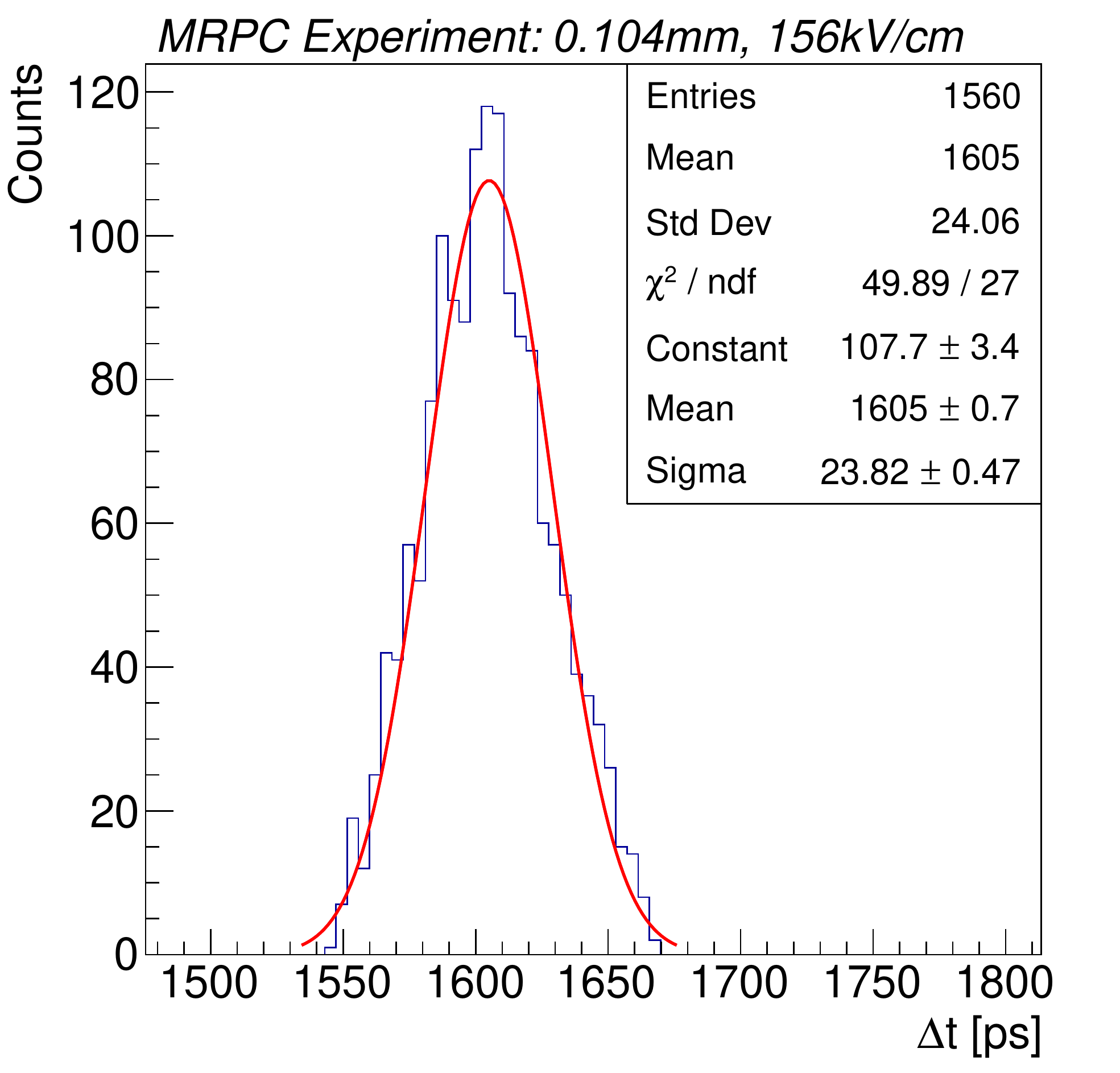}
        \caption{}
        \label{fig:deltadistri}
    \end{subfigure}
    \hspace{1mm}
    \begin{subfigure}[b]{0.48\textwidth}
        \includegraphics[width=\textwidth]{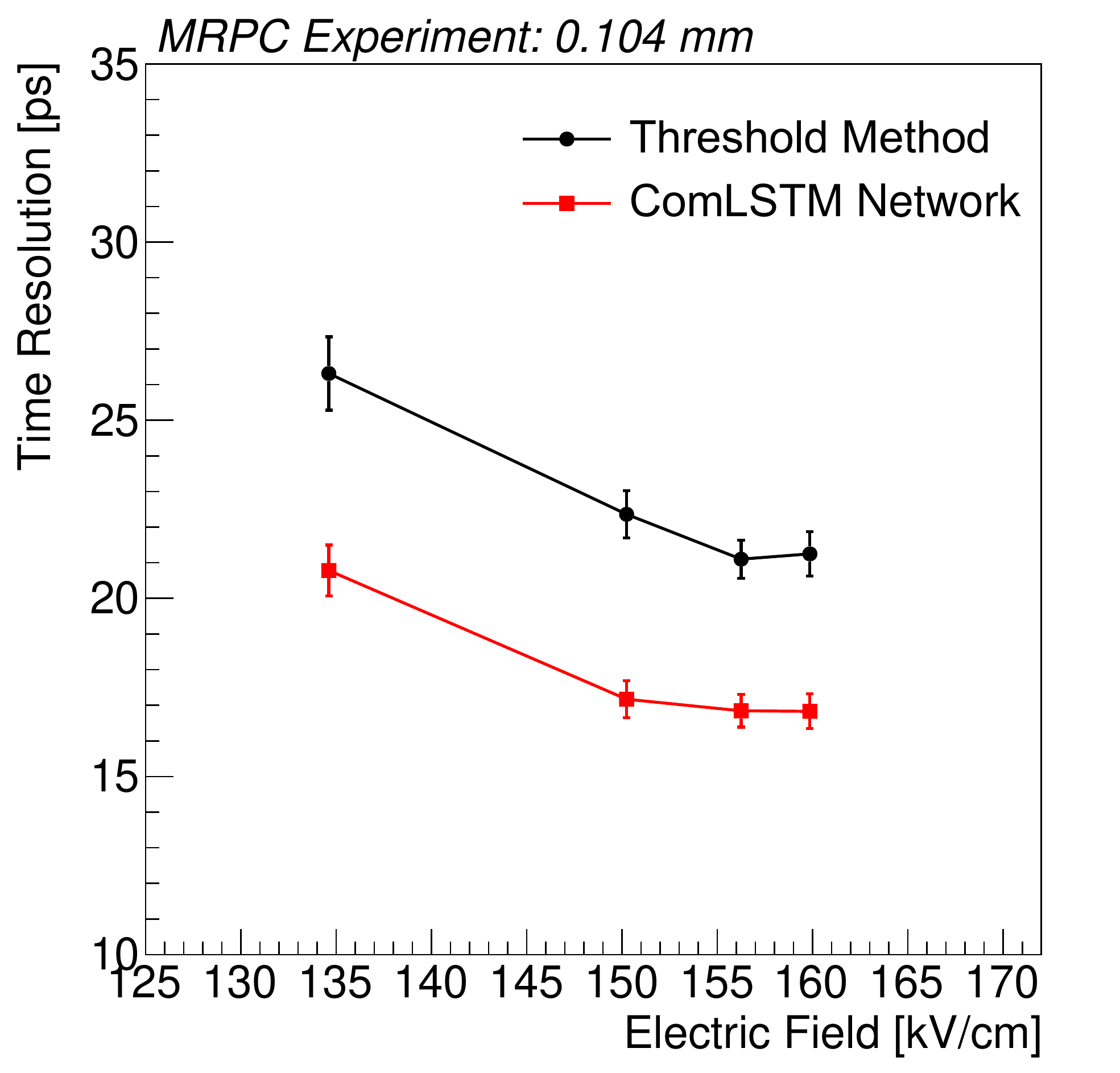}
        \caption{}
        \label{fig:withE}
    \end{subfigure}
    \caption{The time resolution of the MRPC.}
    \label{fig:timereso}
\end{figure}

The black markers and curves in Figure \ref{fig:withE} shows the results given by the threshold method. The sampled signals collected from the left and right sides of the PCB strips on both MRPCs are fit with a 5th polynominal function separately and discriminated by a fixed threshold of 20 fC. Four threshold crossing time $t_{l1}^c$, $t_{r1}^c$, $t_{l2}^c$ and $t_{r2}^c$ are thus obtained. Slewing correction that eliminates the dependence between threshold crossing time and signal amplitude is made. In this case, the time difference between two MRPCs $\Delta t_c$ is calculated according to Eq.\ref{eq:deltat} and corrected with the amplitude of them iteratively. The distribution of $\Delta t_{c}$ after the correction is fit using a gaussian function and the standard deviation over $\sqrt{2}$ is regarded as the time resolution of this method. The time resolution gets improved with respect to the electric field and the best result is around 21.1 ps.

The training data of the ComLSTM is optimized as shown in Section \ref{sec:data}, and the parameters of the FEE are used to simulate the detector working at different electric fields. The rising time $t_r$ of the left and right signals of both the MRPCs are predicted separately by the converged model and are transformed into the first interaction time $t_0$. According to Eq.\ref{eq:deltat}, the difference of the first interaction time between two MRPCs $\Delta t$ is calculated and its distribution at $E=156$ kV/cm is shown in Figure \ref{fig:deltadistri}, where the red curve is a 3$\sigma$ gaussian fit. The time resolution at this condition is $\sigma(\Delta t)/\sqrt{2}=16.84$ ps which well satisfies the requirement of the SoLID experiment. The resolutions given by the ComLSTM at other electric field are shown in Figure \ref{fig:withE} with the red markers and curve. The tendency is consistent with the threshold method but the performance is much better.  

\section{Discussion}
\label{sec:dis}
The ComLSTM neural networks in the previous sections are trained using Monte Carlo simulation data. The advantage of the simulation is that the real time when the particles arrive at the detector is known, so there is a clear correlation between the input (waveform) and the output (rising time) of the network. Models established in this way can be an effective estimator when the simulation data are highly consistent with the experiment. However, this consistency relies not only on an accurate simulation software, but also on a fine calibration of the electronics parameters, which might be time consuming. To avoid the consistency problem, neural network algorithms that based only on the experiment data are also proposed.

In the experiment data, although the true interaction time of every signal is unknown, the time difference between two MRPCs is always constant for any test configuration and vertical particles, and therefore it is used as the labels of the network. The true value of the time difference is defined in two ways: the time in which a vertical incident particle travels through two MRPC at the speed of light $\Delta t_{et1}$, and the time difference of MRPCs reconstructed by the threshold method $\Delta t_{et2}$. The input of the network for every event is a collection of 4 waveforms (left and right of both MRPCS) and their corresponding reference time. The network has the same structure as Figure \ref{fig:comLSTM}. In path $A$ of the ComLSTM, the number of time step of LSTM-420 is 4 which corresponds to 4 different waveforms and the dimension of the input in every time step is 11 which consists of 10 points on the leading edge and a reference time. Path $B$ is the same as Section \ref{sec:meth}. Ideally, $\Delta t_{et1}$ and $\Delta t_{et2}$ should be uniformly distributed within a certain range, which means the distance between two MRPCS is uniformly distributed. However, it is hard to achieve in the experiment. The real cosmic ray tests contain 4 different configurations: MRPCs are placed closely together, or with a spacer in between. The height of the spacer is chosen to be 1, 2, and 4 cm.

\begin{figure}
    \centering
    \begin{subfigure}[b]{0.48\textwidth}
        \includegraphics[width=\textwidth]{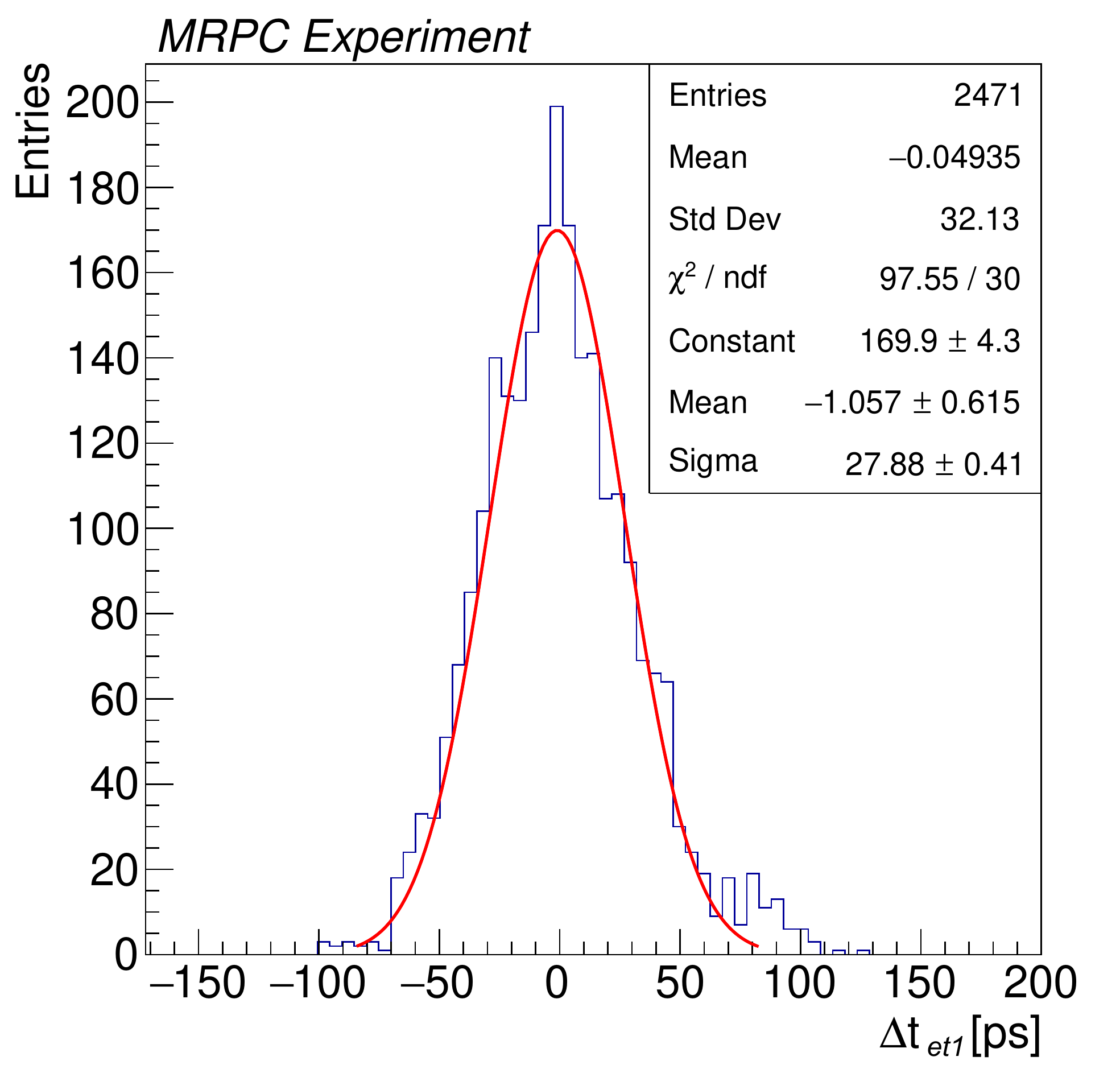}
        \caption{}
        \label{fig:exper1}
    \end{subfigure}
    \hspace{1mm}
    \begin{subfigure}[b]{0.48\textwidth}
        \includegraphics[width=\textwidth]{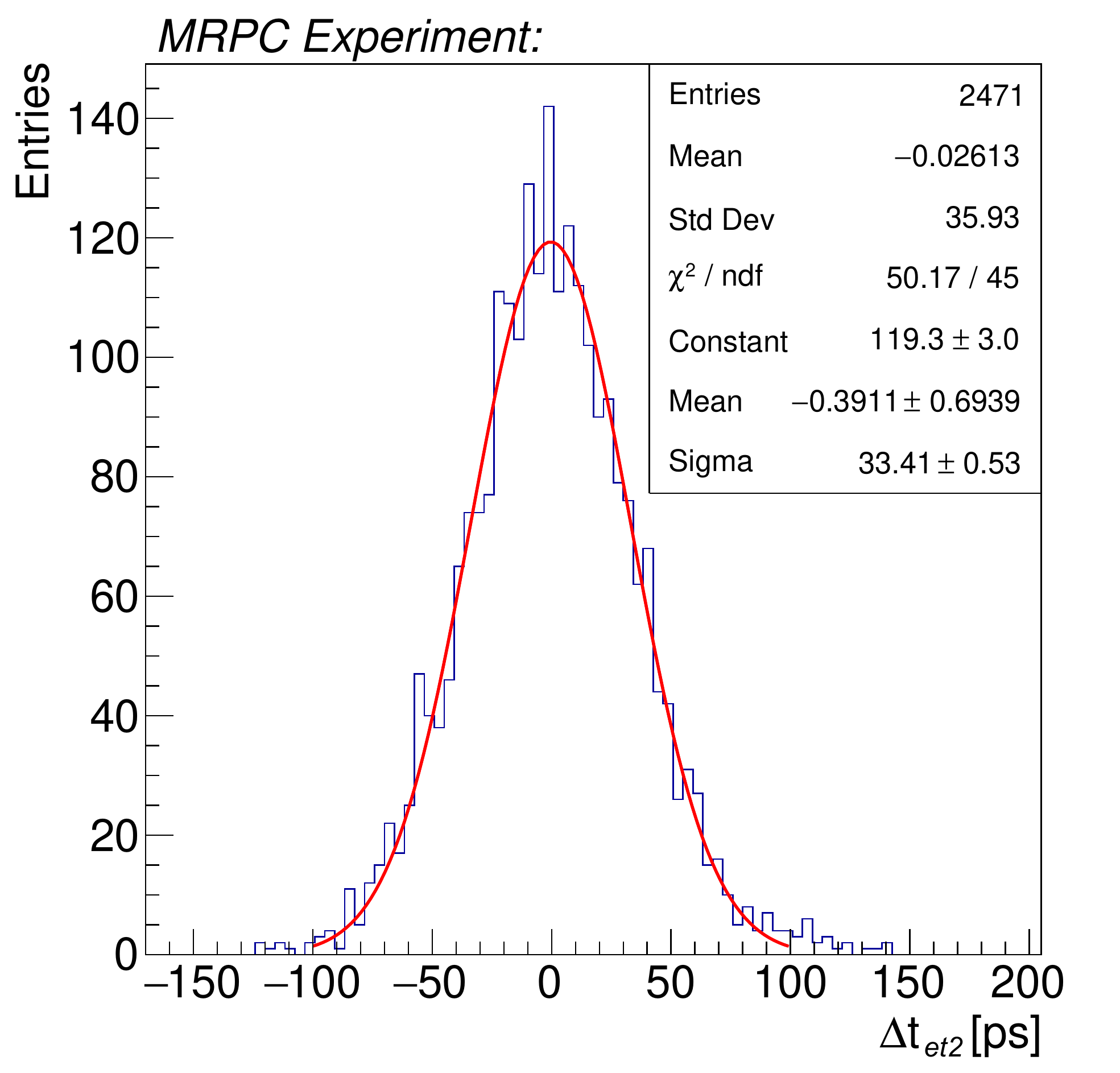}
        \caption{}
        \label{fig:exper2}
    \end{subfigure}
    \caption{The distribution of the time difference predicted by ComLSTM networks with labels defined as $\Delta t_{et1}$ (a) and $\Delta t_{et2}$ (b).}
    \label{fig:exp}
\end{figure}

The experiment data are divided into 2 parts, one for training and the other for testing. Training data are augmented 3 times using label preserving transformations such as choosing the waveforms from the 0th or 2nd point other than the 1st one. Two ComLSTM networks are trained using different labels $\Delta t_{et1}$ and $\Delta t_{et2}$, and finally both of them converge. The testing data are then feed into the network models, and the predicted distributions of the time difference are shown in Figure \ref{fig:exp}, where Figure \ref{fig:exper1} is $\Delta t_{et1}$ and Figure \ref{fig:exper2} is $\Delta t_{et2}$. The distributions are also fit with gaussian functions and according to Eq.\ref{eq:sigmat}, the time resolutions of these two networks are 19.71 ps and 23.62 ps respectively. Both of the results are worse than the simulation based one, because their labels are not as accurate as the simulation. The accuracy of $\Delta t_{et1}$ depends partly on the selection of the vertical incidents, while $\Delta t_{et2}$ largely depends on the accuracy of the threshold method, which means the correlations between the network input (waveforms) and output (time difference) in these two algorithms are less relevant and the features extracted are thus less effective. However, despite of these problems, their resolutions are still at the scale of 20 ps.

\section{Conclusion}
\label{sec:con}
A set of ComLSTM neural networks are proposed and applied to reconstruct the detection time of the MRPC detectors. ComLSTM combines the LSTM and FC neural network and is capable of extracting detailed information from the MRPC signal waveform. The network can be trained with the data from both the Monte carlo simulation and the cosmic ray experiment, while all the results are given using the experiment data. The best time resolution achieved with a thin-gap MRPC which has 4 stacks and 8 gaps per stack is 16.84 ps, much better than the present MRPCs. This result well satisfies the requirements of SoLID experiment and proves the effectiveness and stability of the deep neural network algorithms.


\acknowledgments
The work is supported by National Natural Science Foundation of China under Grant No. 11420101004, 11461141011, 11275108, 11735009. This work is also supported by the Ministry of Science and
Technology under Grant No. 2015CB856905, 2016 YFA0400100.

\bibliography{./ref.bib}   

\begin{thebibliography}{10}
\expandafter\ifx\csname url\endcsname\relax
  \def\url#1{\texttt{#1}}\fi
\expandafter\ifx\csname urlprefix\endcsname\relax\def\urlprefix{URL }\fi
\expandafter\ifx\csname href\endcsname\relax
  \def\href#1#2{#2} \def\path#1{#1}\fi

\bibitem{abelev_performance_2014}
B.~Abelev, A.~Abramyan, J.~Adam, {D. Adamov´a}, M.~Zyzak, Performance of the
  {ALICE} experiment at the {CERN} {LHC}, International Journal of Modern
  Physics A 29~(24) (2014) 1430044.

\bibitem{ackermann_star_2003}
K.~H. Ackermann, N.~Adams, C.~Adler, Z.~Ahammed, {A.N. Zubarev}, {STAR}
  detector overview, Nucl. Inst. Meth. A 499~(2-3) (2003) 624--632.

\bibitem{salabura_probing_2005}
P.~Salabura, G.~Agakichiev, C.~Agodi, et~al., Probing of in-medium hadron
  structure with {HADES}, Nuclear Physics A 749 (2005) 150--159.

\bibitem{besiii_collaboration_construction_2009}
B.~Collaboration, The construction of the {BESIII} experiment,
  Nucl.Instrum.Meth. A598 (2009) 7--11.

\bibitem{solid_collaboration_solid_2017}
S.~Collaboration, {SoLID} {Updated} {Preliminary} {Conceptual} {Design}
  {Report}: {Solenoidal} {Large} {Intensity} {Device} (2017).

\bibitem{akindinov_final_2009}
A.~Akindinov, A.~Alici, P.~Antonioli, et~al., Final test of the {MRPC}
  production for the {ALICE} {TOF} detector, Nucl. Inst. Meth. A 602~(3) (2009)
  709--712.

\bibitem{bonner_single_2003}
B.~Bonner, H.~Chen, G.~Eppley, F.~Geurts, J.~Lamas-Valverde, C.~Li, W.~J.
  Llope, T.~Nussbaum, E.~Platner, J.~Roberts, A single {Time}-of-{Flight} tray
  based on multigap resistive plate chambers for the {STAR} experiment at
  {RHIC}, Nucl. Inst. Meth. A 508~(1) (2003).

\bibitem{wang_detailed_2020}
F.~Wang, D.~Han, Y.~Wang, P.~Lyu, Y.~Li, A detailed study on the intrinsic time
  resolution of the future {MRPC} detector, Nucl. Inst. Meth. A 950 (2020)
  162932.

\bibitem{wang_study_2020}
F.~Wang, D.~Han, Y.~Wang, Y.~Yu, P.~Lyu, B.~Guo, The study of a new time
  reconstruction method for {MRPC} read out by waveform digitizer, Nucl. Inst.
  Meth. A 954 (2020) 161224.

\bibitem{wang_neural_2019}
F.~Wang, D.~Han, Y.~Wang, Y.~Yu, B.~Guo, Y.~Li, A neural network based
  algorithm for {MRPC} time reconstruction, JINST 14~(07) (2019)
  C07006--C07006.

\bibitem{collaboration_neural_2014}
{The ATLAS collaboration}, A neural network clustering algorithm for the
  {ATLAS} silicon pixel detector, JINST 9~(09) (2014) P09009.

\bibitem{aurisano_convolutional_2016}
A.~Aurisano, A.~Radovic, D.~Rocco, A.~Himmel, M.~D. Messier, E.~Niner,
  G.~Pawloski, F.~Psihas, A.~Sousa, P.~Vahle, A {Convolutional} {Neural}
  {Network} {Neutrino} {Event} {Classifier}, JINST 11~(09) (2016) P09001.

\bibitem{wang_standalone_2018}
F.~Wang, D.~Han, Y.~Wang, Y.~Yu, Q.~Zhang, B.~Guo, Y.~Li, A standalone
  simulation framework of the {MRPC} detector read out in waveforms, JINST
  13~(09) (2018) P09007--P09007.

\end{thebibliography}






\end{document}